\documentclass[a4paper,twocolumn,english]{revtex4}
\usepackage[T1]{fontenc}
\usepackage[latin9]{inputenc}
\usepackage{bm}
\usepackage{amsbsy}
\usepackage{graphicx}
\usepackage{amssymb}

\usepackage{graphics}

\usepackage{times}

\usepackage{amstext}

\usepackage{bbm}

\usepackage{babel}

\begin{document}

\title{Indirect Hamiltonian Identification through a small gateway}

\author{Daniel Burgarth$^{1}$}

\author{Koji Maruyama$^{2}$}

\affiliation{$^{1}$IMS and QOLS, Imperial College, London SW7 2BK, UK \\
 $^{2}$Advanced Science Institute, The Institute of Physical and
Chemical Research (RIKEN), Wako-shi, Saitama 351-0198, Japan}
\begin{abstract}
Identifying the nature of interactions in a quantum system is essential
in understanding any physical phenomena. Acquiring information on
the Hamiltonian can be a tough challenge in many-body systems because
it generally requires access to all parts of the system. We show that
if the coupling topology is known, the Hamiltonian identification
is indeed possible indirectly even though only a small gateway to
the system is used. Surprisingly, even a degenerate Hamiltonian can
be estimated by applying an extra field to the gateway.
\end{abstract}
\maketitle

\section{Introduction}

When studying any quantum mechanical system, precise knowledge of
its nature is crucially important. In quantum mechanics, any observable
phenomena can be explained rigorously, in principle, if we have complete
knowledge of the system. More specifically, we need to identify the
states of the system, and the Hamiltonian that governs their dynamics.
Thus, the acquisition of all the relevant information on the states
and Hamiltonian is essential in understanding how nature behaves.
The system of interest may include literally everything quantum mechanical,
from high Tc superconductors to microscopic structures in nanotechnology
or even some highly complex processes in microbiology. 

The full information acquisition is, however, in general very hard
from an operational as well as from a computational and mathematical
point of view, even for small systems \cite{Schirmer2009,Schirmer2008,Young2008}.
For large many-body systems spectroscopy reveals only little information
about the Hamiltonian, and generally local addressing of its components
is required in order to obtain details about the system. Spins which
can be controlled individually operate as a \emph{gateway}, through
which we can access and manipulate the system. A common dilemma is
that such a gateway not only allows us to interact with the system,
but also introduces noise to it. From a Hamiltonian identification
perspective, it is therefore crucial to find minimal gateways that
suffice to obtain \emph{full} knowledge on the system. While this
is impossible to answer for generic systems, bounds can be derived
if the topology of the system is known. In this context, some positive
results have been presented for the case of 1-dimensional (1D) chains
of spin-1/2 particles \cite{Burgarth2009,Franco2009}. That is, the
coupling strengths between neighboring spins can be estimated by accessing
only the spin at the end of the chain. Since schemes to initialize
the state of spins as $|\downarrow\downarrow...\downarrow\rangle$
by operating on the chain end are known \cite{Burgarth2007c}, such
identification of the Hamiltonian is sufficient to determine the dynamics
of the system completely. These results are of interest in their own
right, yet they were limited to the simplest of networks, i.e., 1D
chains. 

In this paper, we suggest an estimation scheme for general graphs
of spins. As well as the details of the Hamiltonian identification
procedure, we give a precise condition for the ``gateway'' (accessible
region) that suffices to make the identification possible. For the
important cases of finite 2D/3D lattices such a gateway is given by
one edge or one face of the lattice, respectively. This is remarkable
because the ratio between the gateway size and the unknown parameters
is much higher than in the 1D case. We will also show that while in
the 1D case the decay properties of the state in the gateway can identify
the Hamiltonian, in the 2D case we need its decay properties as well
as the transport properties within the gateway. Interestingly, our
general condition turns out to coincide with the criterion for the
controllability of spin networks \cite{Burgarth2008}. Our results
here thus indicate that Hamiltonian-identifiable systems are quantum-controllable
and vice versa. Furthermore, they support the physical relevance of
the topological properties discussed later. 

\begin{figure}
\includegraphics[scale=0.7]{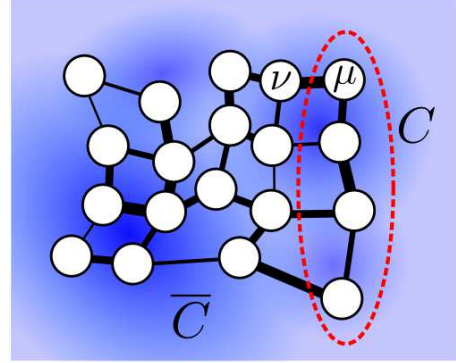}\caption{\emph{\label{fig:2d}All} coupling strengths (black lines) and local
magnetic fields (blue background) of a 2-dimensional network $G=(V,E)$
of spins (white circles) can be estimated \emph{indirectly} by quantum
state tomography on a gateway $C$ (enclosed by the dashed red line).
The coupling strengths and field intensities are represented by the
width of lines and the depth of the background color, respectively.
The labeled spins $\mu$ and $\nu$ are used as examples in the proof
of the main theorem.}

\end{figure}

We will study a network with Heisenberg-type interaction. This allows
us to describe an estimation procedure that is numerically stable,
mathematically simple, and efficient (given that we consider arbitrary
and large systems). What we attempt to estimate are the coupling strengths
between interacting spins and the strengths of local magnetic fields.
Such inhomogeneous fields are very common in experiments, and can
cause much trouble through dephasing. Hence it is worthwhile estimating
them (such analysis was lacking in \cite{Burgarth2009,Franco2009}).
Another interesting new aspect we introduce in this paper is how to
lift degeneracies on the system by applying extra fields on the gateway.
We show that this is always possible, a result which might be relevant
beyond the scope of estimation. 

Our setup is an example of inverse problems that have been actively
studied in plenty of fields in science and engineering. A classical
counterpart among those problems that is closest to our quantum setting
may be the estimation of spring constants in 1D harmonic oscillator
chains \cite{Gladwell2004}. However, the resolution to this (classical)
problem for generic graphs, even the 2D case, is still open. It would
be intriguing if our results in a purely quantum setting could provide
some clues to the analogous problem in classical settings.

\section{Setup and Main Result}

Suppose that we have a network of spin-1/2 particles, such as the
one in Fig \ref{fig:2d}. We assume that we have knowledge of the
graph $G=(V,E)$, which describes the network: nodes $V$ of the graph
correspond to spins and edges $E$ connect spins that are interacting
with each other. The pairwise interaction between spins is Heisenberg
type with a known anisotropy $\Delta,$ and there is an inhomogeneous
magnetic field applied on the spins. Then, the Hamiltonian we consider
has the form 

\begin{eqnarray*}
H & = & \sum_{(m,n)\in E}c_{mn}\left(\sigma_{m}^{x}\sigma_{n}^{x}+\sigma_{m}^{y}\sigma_{n}^{y}+\Delta\sigma_{m}^{z}\sigma_{n}^{z}\right)+\sum_{n\in V}b_{n}\sigma_{n}^{z},\end{eqnarray*}
where $c_{mn}$ represent the \emph{unknown }coupling strengths between
spins $m$ and $n$, and $b_{n}$ the \emph{unknown} intensity of
the magnetic field at $n$, respectively. Here, we also assume $c_{mn}<0$
for all $m$ and $n$, i.e., ferromagnetic interactions, though the
setup is readily generalized to other cases. In the above, $\sigma_{n}^{i}\:(i={x,y,z})$
are the standard Pauli matrices. The purpose of the following will
be to estimate $c_{mn}$ and $b_{n}$ over the entire set $V$ of
spins by only accessing a small gateway, described by a subset $C\subset V$
(See Fig. \ref{fig:2d}). For almost all practical cases of the Hamiltonian
identification problem, analyzing the dynamics in the \emph{single
excitation sector} $\mathcal{H}_{1}$ turns out to be sufficient.
We will thus denote a single excitation state as $|\mathbf{n}\rangle\in\mathcal{H}_{1}$
when the spin $n\in V$ is in the state $|\uparrow\rangle$ and all
others are in $|\downarrow\rangle$ for clarity. The state with all
spins in $|\downarrow\rangle$ will be written as $|\mathbf{0}\rangle.$ 

Naturally, the nice challenge here is to obtain information about
the inaccessible spins $\overline{C}\equiv V\backslash C$, which
could be the large majority of the set. The question is however how
small can the controlled $C$ be such that we can (in principle) still
learn all the couplings and fields in $V?$ Intuitively the knowledge
of the graph structure can be useful for making the estimation efficient.
For instance, the smaller the number of non-vanishing couplings $|E|,$
the more efficiently we can estimate them. However the efficiency
should also depend on the structural property of the graph. 

To answer this question, we need to introduce a property, known as
\emph{infecting,} of a subset $C\subset V$ of the nodes~\cite{Burgarth2007,Burgarth2008,Severini2008,Alon}\emph{.}
In many-body quantum mechanics this property has many interesting
consequences on the controllability and on relaxation properties of
the system~\cite{Burgarth2007,Burgarth2008}. The infection process
can be described as follows. Suppose that a subset $C$ of nodes of
the graph is ``infected'' with some property. This property then spreads,
infecting other nodes, by the following rule: an infected node infects
a ``healthy'' (non-infected) neighbor if and only if it is its \emph{unique}
healthy neighbor. If eventually all nodes are infected, the initial
set $C$ is called \emph{infecting}. The graph in Fig. \ref{fig:2d}
is an example in which $C$ infects $V$ (we encourage the reader
to confirm this by coloring the nodes in region $C$ and applying
the above propagation rule --- this will make the following proof
much more intuitive). With this definition, we can summarize the main
result of the paper as the following
\begin{description}
\item [{Theorem:}] \emph{Assume that that $C$ infects $V.$ Then all $c_{nm}$
and $b_{n}$ can be obtained by acting on $C$ only.} 
\end{description}
This theorem provides an upper bound on the smallest number of spins
we need to access in order to perform Hamiltonian tomography, i.e.
given by the cardinality $|C|$ of the smallest set $C$ that infects
$V.$ To prove the above statement, we first present a lemma and its
proof. 
\begin{description}
\item [{Lemma:}] \emph{Assume that $C$ infects $V$ and that all eigenvalues
$E_{j}$ $(j=1,\ldots,|V|$) in }$\mathcal{H}_{1}$\emph{ are known.
Assume that for all orthonormal eigenstates $|E_{j}\rangle$ in }$\mathcal{H}_{1}$\emph{
the coefficients $\langle\mathbf{n}|E_{j}\rangle$ are known for all
$n\in C.$ Then the $c_{nm}$ and $b_{n}$ are known.}
\end{description}
While the assumptions of the lemma may sound unrealistic, we will
show later how they can be obtained by simple tomography experiments
on $C.$\emph{ }
\begin{description}
\item [{Proof~of~the~Lemma:}]~
\end{description}
We observe that the coupling strengths between spins\emph{ within}
$C$ are easily obtained because of the relation \begin{equation}
c_{mn}=\langle\mathbf{m}|H|\mathbf{n}\rangle=\sum E_{k}\langle\mathbf{m}|E_{k}\rangle\langle E_{k}|\mathbf{n}\rangle,\label{eq:cmn}\end{equation}
where we defined $c_{mm}\equiv\langle\mathbf{m}|H|\mathbf{m}\rangle$
for the diagonal terms. Since $C$ infects $V$ there is a $\mu\in C$
and a $\nu\in\overline{C}\equiv V\backslash C$ such that $\nu$ is
the only neighbor of $\mu$ outside of $C,$ i.e. \begin{equation}
\langle\mathbf{n}|H|\bm{\mu}\rangle=0\;\;\forall n\in\overline{C}\backslash\{\nu\}.\label{eq:null}\end{equation}
For an example see Fig. \ref{fig:2d}. Using the eigenequation, we
obtain for all $j$ \[
E_{j}|E_{j}\rangle=H|E_{j}\rangle=\sum_{m\in C}\langle\mathbf{m}|E_{j}\rangle H|\mathbf{m}\rangle+\sum_{n\in V\backslash C}\langle\mathbf{n}|E_{j}\rangle H|\mathbf{n}\rangle.\]
Multiplying with $\langle\bm{\mu}|$ and using Eq.~(\ref{eq:null})
we obtain\begin{equation}
E_{j}\langle\bm{\mu}|E_{j}\rangle-\sum_{m\in C}c_{\mu m}\langle\mathbf{m}|E_{j}\rangle=c_{\mu\nu}\langle\bm{\nu}|E_{j}\rangle.\label{eq:main4estimation}\end{equation}
By assumption and by Eq.~(\ref{eq:cmn}), the left-hand side (LHS)
is known for all $j.$ This means that up to an unknown constant $c_{\mu\nu}<0$
the expansion of $|\bm{\nu}\rangle$ in the basis $|E_{j}\rangle$
is known. Through normalization of $|\mathbf{\bm{\nu}}\rangle$ we
then obtain $c_{\mu\nu}$ and hence $\langle\bm{\nu}|E_{j}\rangle$.
Redefining $C\Rightarrow C\cup\{\mu\}$, it follows by induction that
all $c_{mn}$ are known. Finally, we have\begin{equation}
c_{mm}=\langle\mathbf{m}|H|\mathbf{m}\rangle=E_{0}-\Delta\sum_{n\in N(m)}c_{mn}+2b_{m},\label{eq:magfield}\end{equation}
where $N(m)$ stands for the (directly connected) neighborhood of
$m,$ and \begin{equation}
E_{0}=\frac{1}{2}\Delta\sum_{(m,n)\in V}c_{mn}-\sum_{n\in V}b_{n}\label{eq:e0}\end{equation}
is the energy of the ground state $|\mathbf{0}\rangle$. Summing Eq.
(\ref{eq:magfield}) over all $m\in V$ and using Eq. (\ref{eq:e0}),
we can have the value of $\sum_{n\in V}b_{n}$, thus that of $E_{0}$
as well, since all other parameters are already known. Then we obtain
the strength of each local magnetic field, $b_{m}$, from Eq. (\ref{eq:magfield}).
$\blacksquare$

\section{Tomography}

Let us now describe how to obtain the information that is assumed
to be known in the lemma. That is, we need to know the energy eigenvalues
$E_{j}$ in $\mathcal{H}_{1}$ and the coefficients $\langle\mathbf{n}|E_{j}\rangle$
for all $n\in C$ by controlling/measuring the spins in $C$. Let
us first consider the case where the eigenvalues in $\mathcal{H}_{1}$
are non-degenerate. The general case will be described in Section~\ref{sec:Degeneracy-and-Efficiency}.
To start the estimation, we initialize the system as $\frac{1}{\sqrt{2}}(|\mathbf{0}\rangle+|\mathbf{1}\rangle).$
As discussed in \cite{Burgarth2007c} this can be done efficiently
by acting on region $C$ only. Then, we perform quantum state tomography
on the spin $n\in C$ after a time lapse $t$. The entire state at
$t$ is now\[
\frac{1}{\sqrt{2}}U(t)(|\mathbf{0}\rangle+|\mathbf{1}\rangle)=\frac{1}{\sqrt{2}}\left(e^{-iE_{0}t}|\mathbf{0}\rangle+\sum_{n=1}^{|V|}f_{n1}(t)|\mathbf{n}\rangle\right),\]
where $f_{n1}=\langle\mathbf{n}|U(t)|\mathbf{1}\rangle$ are the elements
of the time evolution operator in the single excitation subspace.
By repeating the preparation and tomographic measurements on spin
$n$ for various times $t$, we obtain the following matrix elements
of the time evolution operator as a function of $t:$ \begin{equation}
e^{iE_{0}t}\langle\mathbf{n}|U(t)|\mathbf{1}\rangle=\sum_{j}\langle\mathbf{n}|E_{j}\rangle\langle E_{j}|\mathbf{1}\rangle e^{-i(E_{j}-E_{0})t}.\label{eq:fourier}\end{equation}
If we take $n=1$ and Fourier transform Eq.~(\ref{eq:fourier}) we
can get information on the energy spectrum of the Hamiltonian in $\mathcal{H}_{1}$.
Up to an unknown constant $E_{0}$, which will turn out to be irrelevant
later, we learn the values of those $E_{j}$ corresponding to eigenstates
that have non-zero overlap with $|\mathbf{1}\rangle.$ We also obtain
the values of $|\langle\mathbf{1}|E_{j}\rangle|^{2}$ for \emph{all}
eigenstates. Due to the freedom in determining the overall phase of
a state, we can assume that the coefficients for $|\mathbf{1}\rangle$
of all $|E_{j}\rangle$ are real and positive, $\langle\mathbf{1}|E_{j}\rangle>0.$
Hence observing the \emph{decay/revival} of an excitation at $n=1$
we can already learn some $E_{j}$ and all the $\langle\mathbf{1}|E_{j}\rangle$.
This is analogous to the $1D$ case, where this knowledge would suffice
to obtain the full Hamiltonian \cite{Burgarth2009}. 

In arbitrary graphs however this is no longer the case. In fact even
if we observed the decay/revival at each $n\in C$ we would only obtain
the $\left|\langle\mathbf{n}|E_{j}\rangle\right|^{2},$ but could
not determine their phase freely anymore. To obtain the required knowledge
for the Lemma, we need to observe the \emph{transport }within $C.$
This is represented by Fourier transforming Eq. (\ref{eq:fourier})
for $n\neq1,$ allowing us to extract the coefficient $\langle\mathbf{n}|E_{j}\rangle$
correctly, including their relative phase with respect to $\langle\mathbf{1}|E_{j}\rangle$.
We also obtain those eigenvalues $E_{j}$ which have non-zero overlap
with $|\mathbf{n}\rangle.$ Continuing this analysis over all elements
of $C,$ we learn all eigenvalues which have overlap with \emph{some
}$n\in C.$ Could there be eigenstates in $\mathcal{H}_{1}$ which
have no overlap with \emph{any $n\in C$}?\emph{ }The answer is no,
as it is shown in~\cite{Burgarth2007}. Therefore we can conclude
that all eigenvalues in the $\mathcal{H}_{1}$ can be obtained.

Although tomography cannot determine the extra phase shift $E_{0},$
it does not affect the estimation procedure. There are three equations
that seem to require the explicit values of $E_{j}$, namely Eq. (\ref{eq:cmn})
for $c_{mn}$ inside $C,$ Eq. (\ref{eq:main4estimation}) for $c_{\tilde{m}\tilde{n}}$
and coefficients $\langle\tilde{\mathbf{n}}|E_{j}\rangle$ for a spin
outside $C,$ and Eq. (\ref{eq:magfield}) for the magnetic fields.
It is straightforward to see that for $m\neq n$ substituting $E_{j}-E_{0}$
into $E_{j}$ in Eq. (\ref{eq:cmn}) gives the correct $c_{mn}.$
Similarly, the invariance of Eq. (\ref{eq:magfield}) is clear as
it only depends on $E_{j}-E_{0}$. Less obvious is Eq. (\ref{eq:main4estimation}),
however, the key is that the summation over $m\in C$ contains the
diagonal term $c_{\mu\mu}=\langle\boldsymbol{\mu}|H|\mathbf{\boldsymbol{\mu}}\rangle=\sum_{j}E_{j}|\langle\boldsymbol{\mu}|E_{j}\rangle|^{2}.$
Then, by substituting $E_{j}-E_{0}$ into $E_{j}$ in the LHS of Eq.
(\ref{eq:main4estimation}), it is straightforward to confirm that
$E_{0}$ cancels out. Therefore, the precise value of $E_{0}$ is
not necessary for the Hamiltonian identification. Eventually, $E_{0}$
can be calculated by Eq. (\ref{eq:e0}) after having all $c_{mn}$
and $b_{n}$.

\section{Efficiency and Degeneracy\label{sec:Degeneracy-and-Efficiency}}

The efficiency analysis of the Hamiltonian tomography is roughly the
same as in \cite{Burgarth2009}. Due to the conservation of excitations,
the sampling can be restricted to an effective $|V|$- dimensional
Hilbert space, and the speed is some polynomial in $|V|,$ provided
localization is negligible. One difference however is that in arbitrary
graphs it might be less likely that the spectrum is non-degenerate.
An explicit example can be given for a square lattice with equal coupling
strengths, with the spectrum $E_{kj}=E_{k}+E_{j},$ $k,j=1,...,N,$
where the $E_{k}$ the $1D$ energies of the corresponding chain.
A uniform $1D$ system on the other hand would typically be non-degenerate.
Of course \textquotedbl{}exact degeneracy\textquotedbl{} is highly
unlikely; however approximate degeneracy could make the scheme less
efficient. Here, we suggest to lift degeneracies by applying extra
fields on the gateway $C.$ Since $C$ is only a small subset of the
spin, it is not obvious at all that this is possible. We prove the
following perhaps startling property of the infection property:
\begin{description}
\item [{Theorem:}] \emph{Assume that $C$ infects $V.$ Then there exists
an operator $B_{C}$ on $C$ that lifts all degeneracies of $H$ in
the single excitation subspace.}
\item [{Proof:}]~
\end{description}
We will prove the above by explicitly constructing a $B_{C}$ that
does the job. This $B_{C}$ will be very inefficient and even requires
full knowledge of the Hamiltonian, but is only introduced here for
the sake of this proof. Let us denote the eigenvalues of $H$ as $E_{k}$
and the eigenstates as $|E_{k}^{d}\rangle,$ where $d=1,\ldots,D(k)$
is a label for the $D(k)$-fold degenerate states. Let us first concentrate
on one specific eigenspace $\left\{ |E_{k}^{d}\rangle,d=1,\ldots,D(k)\right\} $
corresponding to an eigenvalue $E_{k}.$ Since the eigenstates considered
here are in the single excitation subspace, we can always decompose
them as \begin{equation}
|E_{k}^{d}\rangle_{C\bar{C}}=|\bm{\phi}_{k}^{d}\rangle_{C}\otimes|\bm{0}\rangle_{\bar{C}}+|\bm{0}\rangle_{C}\otimes|\bm{\psi}_{k}^{d}\rangle_{\bar{C}},\label{eq:split}\end{equation}
 where we introduced the unnormalized states $|\bm{\phi}_{k}^{d}\rangle_{C}$
and $|\bm{\psi}_{k}^{d}\rangle_{\bar{C}}$ in the single excitation
subspace of $C$ and $\overline{C},$ respectively. As shown in \cite{Burgarth2007}
we know that $|\bm{\phi}_{k}^{d}\rangle_{\bar{C}}\neq0\:\forall d$.
This is because if there was an eigenstate in the form of $|\bm{0}\rangle_{C}\otimes|\bm{\psi}_{k}^{d}\rangle_{\bar{C}}$
then applying $H$ repeatedly on it will necessarily introduce an
excitation to the region $C,$ in contradiction to being an eigenstate.
In fact the set $\left\{ |\bm{\phi}_{k}^{d}\rangle_{C},\, d=1,\ldots,D(k)\right\} $
must be linearly independent: for, if it was linearly dependent, there
would be complex numbers $\alpha_{kd}$ such that $\sum_{d}\alpha_{kd}|\bm{\phi}_{k}^{d}\rangle_{C}=0,$
and because the eigenstates are degenerate, $\sum_{d}\alpha_{kd}|E_{k}^{d}\rangle_{C\bar{C}}=\sum_{d}\alpha_{kd}|\bm{0}\rangle_{C}\otimes|\bm{\psi}_{k}^{d}\rangle_{\bar{C}}$
would be an eigenstate with no excitation in $C,$ again contradicting
Ref.~\cite{Burgarth2007}. This leads to an interesting observation
that the degeneracy of each eigenspace can be maximally $|C|-$fold,
because there can be only $|C|$ linearly independent vectors at most
in the single excitation sector on $C.$ Thus minimal infecting set
of a graph gives us some bounds on possible degeneracies.

Now we consider a Hermitian perturbation $B_{kC}\otimes\openone_{\bar{C}}$
(to be specified later) on the system and compute the shift in energies.
We shall see that it suffices to assume that $B_{kC}|\mathbf{0}\rangle_{C}=0.$
In first order, we need to compute the eigenvalues of the perturbation
matrix \begin{equation}
_{C\bar{C}}\langle E_{k}^{d}|B_{kC}|E_{k}^{d'}\rangle_{C\bar{C}}=_{C}\langle\bm{\phi}_{k}^{d}|B_{kC}|\bm{\phi}_{k}^{d'}\rangle_{C}.\label{eq:pert_matrix}\end{equation}
 Can we find a $B_{kC}$ such that all eigenvalues differ? For that,
note that $\left\{ |\bm{\phi}_{k}^{d}\rangle_{\bar{C}},\, d=1,\ldots,D(k)\right\} $
are linearly independent, which means that there is a similarity transform
$S_{k}$ (not necessarily unitary, but invertible) such that the vectors
$|\bm{\xi}_{k}^{d}\rangle_{C}\equiv S_{k}^{-1}|\bm{\phi}_{k}^{d}\rangle_{C}$
are orthonormal. The perturbation matrix can then be written as $_{C}\langle\bm{\xi}_{k}^{d}|S_{k}^{\dagger}B_{kC}S_{k}|\bm{\xi}_{k}^{d'}\rangle_{C}.$
If we set \[
S_{k}^{\dagger}B_{kC}S_{k}=\sum_{d}\epsilon_{kd}|\bm{\xi}_{k}^{d}\rangle_{C}\langle\bm{\xi}_{k}^{d}|\]
 we can see that the Hermitian operator $B_{kC}\equiv\sum_{d}\epsilon_{kd}\left(S_{k}^{\dagger}\right)^{-1}|\bm{\xi}_{k}^{d}\rangle_{C}\langle\bm{\xi}_{k}^{d}|S_{k}^{-1}$
gives us energy shifts $\epsilon_{kd}.$ Therefore, as long as we
choose the $\epsilon_{kd}$ mutually different from each other, the
degeneracy in this eigenspace is lifted by $B_{kC}.$ This happens
for an arbitrarily small perturbation $\lambda_{k}.$ We choose $\lambda_{k}$
such that the lifting is large, but in a way such that \emph{no new
degeneracies }are created, i.e. $||\lambda_{k}B_{kC}||<\min\left(\Delta E\right),$
where $\Delta E$ are the energy differences of $H.$ However, the
perturbation $\lambda_{k}B_{kC}$ may well lift other degeneracies
of $H$ ``by mistake''. Note that by construction $B_{kC}$ conserves
the number of excitations in the system (See Eq. (\ref{eq:pert_matrix})).
Therefore, we can now consider the perturbed Hamiltonian $H'=H+\lambda_{k}B_{kC}$
and find its remaining degenerate eigenspaces in $\mathcal{H}_{1}$.
Naturally, the number of degeneracies with $H'$ is less than that
with $H$. Following the above procedure, we pick one eigenspace and
find an operator $B_{k'C}$ that lifts its degeneracy. Keeping $||\lambda_{k'}B_{k'C}||<\mbox{min}(\Delta E')$
we continue to add perturbations, until we end up with a sum of perturbations
that lift \emph{all }degeneracies in $\mathcal{H}_{1}$. $\blacksquare$

The above theorem demonstrates that degeneracies can in principle
be lifted. In practice, we expect that almost all operators will lift
the degeneracy, with a good candidate being an inhomogeneous magnetic
field on $C.$ One could even randomly choose operators on the gateway
$C$ until the system shows no degeneracies. Albeit being inefficient,
our theorem shows that this strategy will eventually succeed. Note
also in the theorem it sufficed to consider operators within $\mathcal{H}_{1}$,
i.e. $B=\sum_{m,n\in C}b_{mn}|\mathbf{m}\rangle\langle\mathbf{\mathbf{n}}|$
with $b_{mn}^{*}=b_{nm},$ so maximally $|C|^{2}$ parameters need
to be tested. For instance, if the system is a chain, $B$ necessarily
corresponds to a magnetic field on spin $1.$

\section{Conclusions}

We have shown how a small gateway can efficiently be used to estimate
a many-body Heisenberg Hamiltonian, given that the topology of the
system is known. It is surprising to see how a simple topological
property of a network of coupled spins - infection - implies so many
far-reaching properties, from control to relaxation, from the structure
of eigenstates to possible degeneracies, and, as we have shown here,
for Hamiltonian identification. 

Our results can be seen as an example of inverse problems in quantum
setting. It would be intriguing to explore a possible link between
ours and similar problems in classical setting, such as 2D graphs
of masses connected with springs. Also, it would be interesting to
study if the methods of \cite{Franco2009}, which does not require
state preparation, can be applied to this setup. A further application
could be found, for example, in estimating the hidden dynamics in
an environment of an controllable system, such as a nanoscale device
\cite{Ashhab2006}. Of course, generalizing the present results to
a wider class of many-body Hamiltonian will be important from both
theoretical and practical perspectives. 
\begin{acknowledgments}
We thank M. B. Plenio and M. Cramer for helpful comments. DB acknowledges
support by the EPSRC grant EP/F043678/1. KM is grateful for the support
by the Incentive Research Grant of RIKEN.
\end{acknowledgments}


\begin{thebibliography}{10}

\bibitem{Schirmer2009}
S. Schirmer and D. Oi, arXiv:0902.3434.

\bibitem{Schirmer2008}
S. Schirmer, D. Oi, and S. Devitt, Institute of Physics: Conferences Series
  {\bf 107},  012011  (2008).

\bibitem{Young2008}
K.~C. Young, M. Sarovar, R. Kosut, and K.~B. Whaley, arXiv:0812.4635.

\bibitem{Burgarth2009}
D. Burgarth, K. Maruyama, and F. Nori, Phys. Rev. A {\bf 79},  020305(R)
  (2009).

\bibitem{Franco2009}
C.~D. Franco, M. Paternostro, and M.~S. Kim, arXiv:0812.3510.

\bibitem{Burgarth2007c}
D. Burgarth and V. Giovannetti,   (2007), proceedings, M. Ericsson and S.
  Montangero (eds.), Pisa, Edizioni della Normale 2008 (arXiv:0710.0302).

\bibitem{Burgarth2008}
D. Burgarth, S. Bose, C. Bruder, and V. Giovannetti, arXiv:0805.3975.

\bibitem{Gladwell2004}
G.~M.~L. Gladwell, {\em Inverse Problems in Vibration} (Kluwer, Dordrecht,
  2004).

\bibitem{Burgarth2007}
D. Burgarth and V. Giovannetti, Phys. Rev. Lett. {\bf 99},  100501  (2007).

\bibitem{Severini2008}
S. Severini, J. Phys. A: Math. Gen. {\bf 41},  482002  (2008).

\bibitem{Alon}
N. Alon, preprint.

\bibitem{Ashhab2006}
S. Ashhab, J.~R. Johansson, and F. Nori, New. J. Phys. {\bf 8},  103  (2006).

\end{thebibliography}
\end{document}